\documentclass[aps,preprint]{revtex4}%
\usepackage{amsmath}
\usepackage{graphicx}%
\usepackage{amsfonts}%
\usepackage{amssymb}
\begin{document}

\preprint{ }
\date{September 5, 2002}
\title{Low field vortex matter in YBCO: an atomic beam magnetic resonance study}
\author{Harald Hauglin}
\affiliation{Department of Physics, University of Oslo, PO Box 1048 Blindern, 0316 Oslo, Norway}
\author{Nathan G. Woodard, Samuel Dapore-Schwartz and Gregory P. Lafyatis}
\affiliation{Department of Physics, The Ohio State University, Columbus, OH 43210-1106}

\begin{abstract}
We report measurements of the low field structure of the magnetic vortex
lattice in an untwinned YBCO single-crystal platelet. Measurements were
carried out using a novel atomic beam magnetic resonance (ABMR)\ technique.
For a $10.7$ G field applied parallel to the \textit{c-}axis of the sample, we
find a triangular lattice with orientational order extending across the entire
sample.We find the triangular lattice to be weakly distorted \ by the
\textit{a-b} anisotropy of the material and measure a distortion factor,
$f=1.16$. Model-experiment comparisons determine a penetration depth,
$\lambda_{ab}=$ 140 ($\pm20$) nm. The paper includes the first detailed
description of the ABMR technique. We discuss both technical details of the
experiment and the modeling used to interpret the measurements.

\end{abstract}
\maketitle

\section{Introduction}

Considerable theoretical and experimental effort has been devoted to better
understanding magnetic flux vortices and magnetic vortex /flux-line lattices
in superconductors. This activity has been driven by physical phenomena that
have been predicted and observed in high-T$_{C}$ materials and the practical
importance of the role the vortex lattice plays in achieving large critical
currents in the presence of magnetic fields. Existing techniques for imaging
vortices provide largely complementary information. Bitter decoration can
produce a one-time map of vortex locations for many types of material in
fields of up to a few hundred Gauss, provided the vortices do not
move\cite{Gammel87}. Scanning tunneling microscopy can show the structure of
individual vortices and the vortex lattice but is limited to studying
atomically flat samples with very clean surfaces\cite{Maggio-Aprile95,Hess89}.
Electron holography can image the structure and motion of the flux-line
lattice in small fields, also for special, thin, flat samples\cite{Harada93}.
Real time magneto-optical vortex imaging\cite{Goa01}, has so far only been
demonstrated in conventional type II superconductors in weak applied fields.
Scanning Hall Probe Microscopy can image quasi-static vortex structure in weak
fields\cite{Oral98}. Bending \cite{Bending99} \ has recently reviewed local
probes of vortices. Small angle neutron scattering can yield precise
information about the lattice structure in the\emph{\ bulk} of a sample but
often requires long integration times and large sample volumes
\cite{ForganPaulMook90}.

In a recent paper\cite{I}, we demonstrated for the first time a novel ``Atomic
Beam Magnetic Resonance'' technique (abbreviated, ``ABMR,'' below) for
studying magnetic vortices and flux-line lattices. Very recently, we have used
this technique to study the vortex matter phase diagram of YBCO very near
$T_{C}$ .\cite{nature} The basic idea --- illustrated in Fig. 1 --- is to
allow an atomic beam to skim across the surface of a superconducting sample
and measure the rate that rf magnetic resonance (hyperfine) transitions are
excited in atoms as they pass over the sample's flux-line lattice. Transitions
are resonantly driven in atoms where the atom's velocity and the spacings of
vortices along its path combine to make an oscillating magnetic field
component at the magnetic resonance frequency. Experimentally, we measure the
excitation probability for atoms \emph{as a function of their velocity} and
work backwards to infer spatial characteristics of the flux-line lattice.
Below, we show that, in the weak excitation limit, our measured signal is
proportional to the spatial Fourier transform of lattice's autocorrelation function.

\emph{This }paper presents and discusses new ABMR\ measurements made on vortex
lattices in a detwinned single-crystal YBCO sample. In addition, this paper
presents, for the first time, details of the ABMR experimental technique: we
discuss technical aspects of the apparatus and develop the theory and modeling
used to interpret results. We begin in Section 2 with an overview. The aim
here is to lay out, succinctly, the \emph{essential} physics of the
measurements in the context of studying vortex lattices. The atomic physics
details needed for accurate quantitative modeling are left for later sections.
We show representative experimental data, and draw conclusions immediately,
\emph{without} sophisticated modeling. In section 3 we describe the apparatus
and discuss technical details that may affect the measurements. Section 4
contains the theoretical basis of the technique and derives a two-level Master
equation model that we have found invaluable in interpreting experimental
results. Section 5 compares the experimental data with model predictions to
develop a \emph{detailed} description of the vortex lattice.

\section{Overview}

The idea behind our experimental approach was originally suggested by Brown
and King in the early 1970s. \cite{Brown71} \ We skim a thermal beam of atomic
potassium, $\ \ $mostly$\ ^{39}K$, across the vortex lattice of a
superconducting sample. The lowest electronic state of $^{39}K$ , the $4s$
state (Fig. 2) has two hyperfine levels $F=1$ and $F=2$ separated by $\Delta
E/h\equiv f_{0}=462$ MHz. An oscillating magnetic field at that frequency will
resonantly drive transitions between these two levels. A $^{39}K$ atom in the
$\ F=1$ level, travelling through the \emph{spatially inhomogeneous} magnetic
field just above the surface of a type II superconductor in the mixed state
sees a \emph{time-dependent} magnetic field that depends on the atom's
velocity and the spatial pattern of vortices in the superconductor. If the
frequency spectrum of that field has a component at 462 MHz, transitions to
the $F=2$ level will be resonantly driven. Atoms with different velocities are
sensitive to different \emph{spatial} periodicities of the vortex lattice
magnetic field. As an example, an atom with a velocity of 462 m/s --- a
typical velocity for an atom in a thermal potassium beam --- will be excited
by vortices spaced 1 $\mu$m apart, the typical nearest neighbor distance for
the vortex lattice with a flux density of about 20 G. \emph{Experimentally},
we measure the excitation probability for atoms as a function of their
velocity and use that measurement to identify important length-scales in the
vortex lattice.

The essential experimental operations, then, are: (1) preparing atoms in the
initial hyperfine level; (2) passing atoms over the vortex lattice of a
superconducting sample, and; (3) detecting the fraction of atoms excited as a
function of velocity.

We carry out these out as follows (see Fig. 3): (1) The atomic beam is
prepared by optically pumping all atoms to the $F=1$ level of the electronic
ground state --- the initial level of the 462 MHz magnetic resonance
transition. \ (2) 1 meter downstream, atoms pass over the surface of a
superconductor where vortices may drive the 462 MHz magnetic resonance
transition. Atoms travel along the superconductor's surface for a distance of
about a millimeter. Our modelling indicates that only those passing within 1
$\mu$m of the surface are significantly excited. (3) Atoms excited by the
vortex lattice are detected using laser induced fluorescence from an optical
transition out of the $F=2$ ``final'' hyperfine level. The Doppler shift
between the atomic optical transition and the laser tuning allows a specific
velocity-class of atoms to be excited and a data run consists of progressively
tuning into resonance and measuring fluorescence for the different
velocity-classes of atoms in the atomic beam.

Fig. 4 and Fig. 5 show representative data taken for vortex lattices created
in a single-crystal detwinned YBCO platelet. \ The sample was cooled through
its superconducting transition in the presence of a weak bias field
(``field-cooled''). For these data, the direction of the bias field is
perpendicular to the superconductor. This is the YBCO sample's crystalline
$c$-axis and the $z$-axis in the analysis below. The atomic beam passes over
the sample parallel to the $a$-axis of the crystal, our ``$x$-axis.'' To
identify important lengths in the vortex lattice, we display the measured
excitation probability as a function of the length scale,
\begin{equation}
l=\frac{v}{f_{0}}, \label{resonance condition}%
\end{equation}
for which a given atom velocity, $v$, is sensitive. In Fig. 4 we show a series
of measurements taken for different bias fields and observe the expected
qualitative trend --- as the field is increased, the vortices become more
tightly packed.

Sometimes, \emph{structural }information may be immediately extracted from
measurements. Fig. 5 shows data taken at 10 K after \textquotedblleft field
cooling\textquotedblright\ the sample in a 12 G bias field. Below, we show how
these data, through comparisons with model predictions, can provide a detailed
picture of the vortex lattice. Here, we will take a less formal look at the
measurements. Decoration experiments by Dolan et al \cite{Dolan89} have found
that triangular vortex lattices form in YBCO samples with the lattice slightly
distorted by being compressed along the a-axis of the crystal. In Fig. 6, we
show an undistorted triangular lattice and consider first the ABMR signal that
would be expected from it. Now, the strongest peaks in our data are due to
periodically spaced rows of nearest neighbor atoms. There are three sets of
such rows, as are indicated on the figure. We do not know, \emph{a priori},
the orientation of the vortex lattice with respect to the atomic beam. If they
were aligned as shown in Fig. 6A), \ we would expect a single peak --- the
periodicity along the atomic beam for the dashed rows is the same as for the
dotted rows and the beam travels along those rows indicated by the solid line.
Thus, the fact that we see\emph{\ two} peaks in Fig. 5 indicates that the
vortex lattice's orientation is \emph{tilted} with respect to the atomic beam.
And the fact that we see \emph{only} two peaks suggests that vortex lattice
has the \emph{same orientation} over the entire surface of the sample
\cite{Note1} Fig. 6B)\ considers this case and the darker bars indicate the
principal periodicities seen along the atomic beam. For an angle of tilt,
$\theta,$ these are found to be:%

\begin{equation}
d_{1}=\frac{\sin60^{0}}{\sin(60^{0}-\theta)}d_{0}\hspace{0.5in}d_{2}%
=\frac{\sin60^{0}}{\sin(60^{0}+\theta)}d_{0}\hspace{0.5in}d_{3}=\frac{\sin
60^{0}}{\sin\theta}d_{0} \label{peaks}%
\end{equation}
Where $d_{0}$ is the nearest neighbor vortex spacing. The \textit{ratios }of
these may be used to find the tilt between the atomic beam and the vortex
lattice: for a given tilt, the ratios are independent of the vortex density
(i.e. magnetic field) and are unchanged even if the vortex lattice is
compressed along the axis of the beam:%

\[
\frac{d_{1}}{d_{2}}=\frac{\sin(60^{0}+\theta)}{\sin(60^{0}-\theta)}
\]
Identifying the 1.68 $\mu m$ and 1.24 $\mu$m peaks in Fig 6 with $d_{1}$ and
$d_{2}$ , respectively, determines that the vortex lattice is tilted 15$^{0}$
from the atomic beam --- the crystalline $a-$axis. The relative orientation of
the vortex lattice and the underlying crystalline axes was found to be a
robust property: we have made vortex lattices many times in this sample and
found the lattice to form \emph{\ always }oriented the same way.
Interestingly, there is \emph{no} immediately obvious connection between the
orientation of the vortex lattice and either the underlying crystalline
symmetry or the edges of the sample.

The positions of the peaks in Fig. 5 together with the vortex density can be
used to determine the \emph{distortion} of the vortex lattice relative to the
atomic beam (= $a$-axis of the crystalline host in the present case). For an
undistorted triangular lattice, the nearest neighbor spacing is $d_{0}%
=\sqrt{2/\sqrt{3}}\sqrt{\phi_{0}/B},$where $\phi_{0}$ is the quantum of flux
and $B$ is the flux density perpendicular to the surface. Using a miniature
Hall \ probe array we measured the mean field at the surface of the sample and
found it to be 10.7 G --- uncertainty due to to screening by the sample makes
this measurement necessary. Now, Eq. \ref{peaks} predicts that an
\emph{un}distorted 10.7 G vortex lattice tilted $\theta=$15$^{0}$ with respect
to the atomic beam, should produce peaks at $d_{1}^{\prime}=1.83$ $\mu$m and
$d_{2}^{\prime}=1.34$ $\mu$m. However, we measure $d_{1}=$1.68 $\mu$m and
$d_{2}=$1.24 $\mu$m. Therefore, the vortex lattice in the sample must be
compressed along the beam axis (the crystalline $a-$axis) by $d_{1}^{\prime
}/d_{1}=d_{2}^{\prime}/d_{2}=1.08$ and, to produce the correct vortex density,
it must be stretched along the $b$-axis by a corresponding amount. Dolan et
al\cite{Dolan89} define the distortion factor, $f$ , as the $b$-axis
scaling/$a$-axis scaling. In our case, $\ f=1.08^{2}=1.16.$ This value is
consistent with the measurements of Dolan et al\cite{Dolan89}, who observed
distortions ranging from $f=1.11$ to $f=1.15$ in the samples they decorated.
More recently, the small angle neutron scattering studies of Johnson et al
\cite{JFL99} found $\ f=1.18$. This value was later corroborated by muon-spin
measurements on the same sample that found $f=1.16$.\cite{AOL00} Both Dolan et
al and Johnson et al argue that the distortion factor is equal to the ratio of
the penetration depths along the crystalline axes, $f=\gamma_{ab}\equiv
\lambda_{a}/\lambda_{b}$ , and both papers contain discussions that compare
with experiments that otherwise measure those penetration depths. Briefly, our
value, $f=1.16$, is somewhat lower than the $\gamma_{ab}=$ 1.37 to 1.6
reported in polarized reflectivity measurements \cite{WTR98} and at the low
end \ of the range, 1.2 to 1.8 found in the Josephson tunneling studies of Sun
et al.\cite{SHK95} Finally, we note \ that the third peak predicted by Eq.
\ref{peaks} for the lattice in Fig 5, $\ d_{3}$ $\thicksim$4.5 $\mu$m, is out
of the range of sensitivity of the present experiment.

While crystalline vortex lattices with sample-wide order provide the richest
data, \emph{any }vortex arrangement will generate a signal. Fig. 7, for
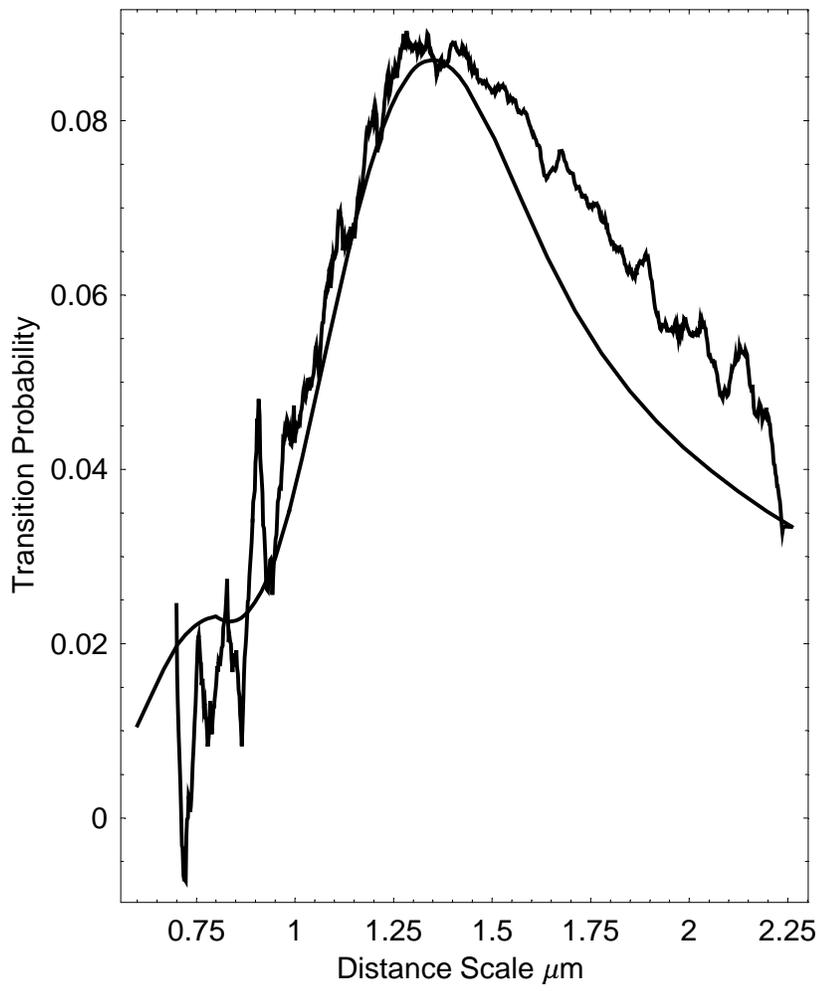
example, shows data acquired from the vortex lattice of a 100 nm thick niobium
film in a 13.4 G magnetic field applied perpendicular to the film's surface.
The single asymmetric peak --- steep on the short distance side, gently
sloping on the long distance side --- is characteristic of a strongly
disordered vortex lattice with only short range translational correlations.
These data are analogous to the X-ray diffraction of an amorphous solid. The
model-generated curve superimposed on the experimental data is for a vortex
lattice with a translational correlation length about 4 times the nearest
neighbor spacing\cite{I}. We discuss details of the modeling below. For now,
Fig. 8 concludes this overview section with a gallery of model predictions for
several different phases of vortex matter.

\section{Experimental details}

In this section, we describe the apparatus and discuss technical details
important to the measurement. See Fig. 2 and Fig. 3. Initially, a thermal
potassium beam is produced in an oven operated at about 400$^{\circ}$ C. The
thermal velocity distribution of the atoms in the beam provides sufficient
quantities of atoms to allow measurements for velocities between 200 and 1000
m/s . For the 462 MHz hyperfine transition, these velocities probe distances
$\sim$0.5 --- 2.5 $\mu$m. A triangular lattice with 0.5 $\mu$m nearest
neighbor spacings corresponds to an 80 G applied field. Thus for studying
superconductors, this represents a ``low field'' diagnostic tool.

\subsection{State Preparation}

The ground state, $4s^{\quad2}S_{1/2}$ , hyperfine levels in the thermal
atomic beam coming from the oven will be statistically populated. A laser
tuned to the 770 nm, $4s^{\quad2}S_{1/2}(F=2)\rightarrow4p^{\quad2}%
P_{1/2}(F=2),$ transition \ optically pumps almost all of the atoms to the
lower ($F=1$) hyperfine level. A pseudo depolarizer in the pumping laser beam
allows all magnetic states of the $4s^{\quad2}S_{1/2}(F=2)$ atoms to be moved
to the lower level. Typically \ 99.5\% of the atoms wind up in that level.

\bigskip

\subsection{Sample Region}

The atoms next pass to a differentially pumped sample chamber. For this work,
the sample was a twin-free, single-crystal YBCO platelet with dimensions 0.7
mm $\times$ 1.7 mm $\times$ 0.1 mm. It was grown by a self-flux
method\cite{Liang92} at the Ohio State University. It has a sharp
superconducting transition ($\Delta T=0.3$ K) with an onset at $T_{C}=93.0$ K.
The sample was thermo-mechanically detwinned in an oxygen atmosphere using a
platinum anvil similar to the apparatus in ref. \cite{Giapintzakis89}.

The sample was mounted to a copper stage that in turn was attached to an LHe
reservoir via a thermally resistive stainless steel link. The stage's
temperature was measured with a silicon thermometer and could be varied from 5
K to
$>$%
100 K using a resistive heater. To reject those atoms that pass too far from
the sample to contribute to the signal but close enough to contribute to the
background in the detector, a tunnel shaped fixture was mounted above the
sample that served as an aperture for the atomic beam. This aperture extended
the length of the sample along the beam and allowed through only those atoms
passing within 2 $\mu m$ of the sample's surface. Low resistance,
$R<$0.1$\Omega,$ leads were mounted to the sample in a 4-wire configuration so
that we could measure the sample's resistance and also drive transport
currents through the sample perpendicular to the atomic beam. Three pairs of
coils mounted outside of the vacuum system allow us to apply small fields to
samples in arbitrary directions.

Immediately following the sample chamber is a Stern-Gerlach magnet that serves
the role of a beam stop and dumps \emph{un}excited atoms from the beam --- the
``main beam;'' This step was included because, otherwise, off-resonance
fluorescence from the unexcited atoms caused large backgrounds in the
detector. Two side-effects of the Stern-Gerlach magnet are 1) in addition to
filtering all out atoms in the $F=1$, lower level, it also removes from the
atomic beam ``signal'' atoms excited to the $F=2,M_{F}=-2$ \ state and 2) the
direction of atoms exiting the magnet depend on their velocity. To send
different velocity classes of atoms on to the detector, it is necessary to
vary the strength of the Stern-Gerlach magnet.

\subsection{Detector}

Experimental difficulties in detecting the small numbers of
vortex-lattice-excited atoms as a function of velocity lead to an involved
detection scheme. At the heart of the detection system is the ``detection
laser'' that drives resonance fluorescence in atoms that were excited by the
vortex lattice. The detection laser is directed nearly antiparallel to the
atomic beam and for a given frequency, it excites a \ velocity class of atoms
with $\delta v\sim20$ m/s. A portion of the fluorescence is collected by an
optical system and detected by a high efficiency detector. In our earlier
work, \cite{I} this detector was a photomultiplier tube. For our more recent
work we use a large-area avalanche photodiode that is cooled with liquid
nitrogen\cite{Hofstedler}. There is a magnetic field parallel with the laser
at the beams' intersection and the laser light is $\sigma^{+}$ polarized. An
important feature of this arrangement is that, while atoms may be excited by
the vortex lattice to any of the Zeeman states in the of upper level of 462
MHz transition of, the detection system is especially sensitive to those atoms
that end up in the $F=2$, $M_{F}=+2$ state . An atom in this state is driven
by the laser in a cycling transition: the laser light can excite only to the
$M_{F}^{\prime}=+3$ of the upper level of the optical transition and this
level can decay only back to the $M_{F}=+2$ state where it may be reexcited,
repeatedly. Typically an atom will produce on the order of 200 fluorescence
photons of which 40 will be collected and detected.

We found that, despite extensive baffling, the ``detection laser'' caused
large fluctuating scattered light backgrounds in the detector. Phase sensitive
(lockin) detection can discriminate against such backgrounds but this requires
a modulation of the atomic beam signal. For this reason, a second beam from
the 770 nm optical pumping laser was directed across the atomic beam just
after the Stern-Gerlach magnet. This second pumping laser beam returns excited
atoms to the lower level of the 462 MHz transition thus turns off the signal
at the detector. We mechanically chopped the second pumping laser beam at a
frequency of 140 Hz. This suitably chopped the signal at the detector and
allowed us to carry out lockin detection at that frequency.

Since our initial work \cite{I}, we have added a stage that transfers
vortex-lattice-excited (signal) atoms from other $F=2$ magnetic substates into
the $M_{F}=+2$ state for which the detection system is most sensitive. Doing
this both\ increases the signal and provides better defined the experimental
conditions for quantitatively interpreting results. The latter is because the
$M_{F}$ states individually fluoresce at slightly different frequencies.
Without this step, it is conceivable that structure in the detected signal
could be due to the Zeeman structure of the detection transition and
\emph{not} structure in the vortex lattice. In addition, knowing that all
vortex-lattice-excited atoms contribute identically to the fluorescence signal
considerably simplifies the quantitative analysis and modeling of the
experiment. To implement this operation, just upstream of the detector, see
Fig. 3, the atomic beam is crossed transversely by a beam picked off the
detection laser. That beam is $\sigma^{+}$ polarized. A weak ($\sim2$ G)
magnetic field is oriented along the laser beam to define a quantization axis.
As atoms scatter photons from this laser, they are moved toward higher $M_{F}$
states with the net result that $>90\%$ of the vortex lattice excited atoms in
$M_{F}=-1,0,1$ Zeeman states are moved to the $M_{F}=2$ state and detected.

\bigskip

\subsection{Data Acquisition}

Data acquisition consists of measuring fluorescence as a function of velocity
\ for the atoms excited by the vortex lattice. Two parts of the detection
scheme have velocity sensitivity: the Stern-Gerlach magnet's steering and the
Doppler shifted resonance frequency of the atoms' optical transition. To take
data, these must be changed synchronously. To make the laser resonant
\emph{with a particular velocity class} of atoms, it is most convenient to
keep the laser itself locked to a single frequency and to tune the atomic
transition into resonance using the Zeeman effect. The detector magnet
provides the required field. Data are acquired by changing in step the
currents to the Stern-Gerlach magnet and the detector magnet while measuring
the avalanche photodiode current using a lockin amplifier that is referenced
to the chopped second pumping beam. Typically, a data set consists of
fluorescence recorded for 1000 points (velocities) and takes 30 seconds to acquire.

To measure the excitation \emph{probability} as a function of velocity, we
proceed as follows. (1) \emph{\ }The sample is warmed to a temperature above
$T_{c}$ where there is no vortex lattice and data are recorded. This gives a
warm pumped data set, WP, and shows how effective the initial state
preparation is. (2) The sample is then allowed to cool in an applied magnetic
field. For each temperature of interest, we record two fluorescence
distributions: \emph{\ }The raw `cold pumped,' CP signal, and the `cold
unpumped,' U signal that we get by blocking the first pumping beam. The latter
serves as a reference for the measurement, since the strength of the
fluorescence signal in this case corresponds to the intensity of the atomic
beam coming from the oven.

The signals are smoothed by performing a 20 point running average of the raw
data and are reduced by subtracting the warm pumped data WP from the cold
pumped signal CP and dividing by the unpumped distribution U. This procedure
is shown in Fig. 9 for a representative set of data. Note that this way of
normalizing the experimental magnetic resonance profiles automatically
measures the \emph{absolute} excitation probability caused by the vortex lattice.

\section{Modeling}

We interpret our experimental results by comparing our measurements with
theoretical signals predicted for likely vortex lattices. To predict signals
we require two theoretical inputs. First we need the relation between the
structure of a vortex lattice in a sample and the magnetic field above the
sample's surface; Second, we need to determine the excitation of a atom due to
the time-dependent magnetic field it sees as it passes through the field of
the vortex lattice.

\bigskip

\subsection{\bigskip The field of a vortex lattice}

The magnetic field above the sample surface may be found by solving the
London-Maxwell equations with the appropriate boundary conditions. Marchetti
\cite{Marchetti92} found that for a bias field applied parallel to the major
anisotropy ($\widehat{\mathbf{c}}$) axis of a sample, the partial spatial
Fourier transform --- over $x$ and $y$ \ directions --- of the magnetic field
at a distance $z$ above the sample surface is given by
\begin{equation}
\mathbf{B(q},z)=\left[  \frac{\phi_{0}}{\lambda_{\text{ab}}^{2}}%
\frac{(\mathbf{\hat{z}}-i\mathbf{\hat{q}})e^{-qz}}{\alpha(\alpha+q)}\right]
\sum_{l}e^{i\mathbf{q\cdot R}_{l}}, \label{fieldstructure}%
\end{equation}
where
\begin{equation}
\alpha=\sqrt{q_{x}^{2}+q_{y}^{2}+\frac{1}{\lambda_{ab}^{2}},}%
\end{equation}
$\{\mathbf{R}_{j}\}$ are the positions of the vortices at the sample's
surface, $\lambda_{a},\lambda_{b}$, $\lambda_{c},$ are the magnetic
penetration depths along the crystalline a-, b- and c-axis, $\lambda
_{ab}=\sqrt{\lambda_{a}\lambda_{b}}$ , $\gamma=\lambda_{c}/\lambda_{ab},$
$\mathbf{q=(}q_{x},q_{y})$, $\phi_{0}=20.7$ G$\left(  \mu\text{m}\right)  ^{2}
$is the flux quantum, and $\mathbf{\hat{z}}$ and $\mathbf{\hat{q}}$ are unit vectors.

\bigskip

\subsection{Excitation of Atoms by the vortex field}

Next, we the consider the excitation of a potassium atom by the \ fluctuating
field, $\mathbf{B}(t)$, \ in its rest frame as it passes over a vortex
lattice. Relevant atomic structure is shown in Fig. 2. Initially, the atomic
beam is optically pumped to the $F=1$ level and we expect that the magnetic
states of this level will be equally populated at the beginning of a
measurement. We consider the excitation of atoms to magnetic states of the
$F=2$ level. The interaction Hamiltonian of an atom in a time varying magnetic
field is:
\begin{equation}
H^{\prime}(t)=-\mathbf{\mu\cdot B}(t)
\end{equation}
here $\mathbf{\mu}$ is the magnetic dipole moment operator of the atom. For a
ground state potassium atom, the electron has no orbital angular momentum and
the magnetic moment of the nucleus is negligibly small, so the magnetic dipole
moment operator is just that of the valence electron's spin: $\mathbf{\ \mu
=-2}\mu_{B}\mathbf{S}$; where $\mu_{B}$ is the Bohr magneton and $\mathbf{S}$
is the dimensionless electron spin operator. Expressing the magnetic field in
terms of its spherical components: $B_{\pm1}=(B_{x}\pm iB_{y})/\sqrt{2}$ and
$B_{0}=B_{z}$, gives:
\begin{equation}
H^{\prime}(t)=\mathbf{2}\mu_{B}\sum_{j=-1,0,1}(-1)^{j}S_{j}B_{-j}(t)
\label{interaction hamiltonian}%
\end{equation}
The general problem of the effect of an arbitrarily varying magnetic field on
an arbitrary mixture of states in the $4s$ manifold is difficult. We have
carried out a limited number of calculations by solving the full
time-dependent Schr\"{o}dinger equation using this Hamiltonian in which
$\mathbf{B}(t)$ is found by following specific paths over a candidate vortex
lattices. These calculations are extremely time consuming and generally
obscure the physics essential to the excitation process. These calculations
are carried out to check the reliability of the approach that we usually use
and which is described next.

We begin by using first order perturbation theory to find the sample-averaged
excitation rate of the $F^{\prime\prime}=1\rightarrow F^{\prime}=2,$
hyperfine\ transition for an atom with a given velocity and height traveling
above the superconductor. Here, the convention is that double primed variables
refer to the energetically lower state of a transition and single primed
variables refer to the energetically higher state. We take as a basis the
hyperfine (variables=electronic spin, nuclear spin) energy eigenstates for the
mean field \ that we measure at the sample's surface. The quantization axis is
given by this field --- i.e. is perpendicular to the sample surface. The
magnetic resonance transition is a magnetic dipole transition with selection
rules $\Delta M_{F}=\pm1,0$ and thus within first order perturbation theory,
the problem of excitation by the vortex lattice of the initial, $F^{\prime
\prime}=1$ level, reduces to nine uncoupled two-state problems corresponding
to the nine allowed $M_{F}^{\prime\prime},F^{\prime\prime}=1\rightarrow
M_{F,}^{\prime}F^{\prime}=2$ transitions.

In Appendix 1, we derive the first order perturbation result for the
excitation rate between specific Zeeman states. For atoms with velocity, $v$,
traveling a height, \ $z$, above the sample, $R_{ge}(v,z)$, the rate of
excitation from the Zeeman state, $g,$in the ground state manifold, to $e$ in
the excited state manifold is given by:%

\begin{equation}
R_{ge}(v,z)=2\eta_{j}\left|  M_{eg}\right|  ^{2}\frac{1}{\pi v}\frac{B\phi
_{0}}{\lambda_{\text{ab}}^{4}}\left(  \frac{\mu_{B}}{\hbar}\right)  ^{2}%
\int\limits_{-\infty}^{+\infty}dq_{y}\frac{e^{-2qz}}{[\alpha(\alpha+q)]^{2}%
}S_{2}(\mathbf{q}), \label{rate coefficient}%
\end{equation}

Where $\eta_{j}=1$ for $\Delta M_{F}=0$ transitions and $\eta_{j}=\frac{1}{2}$
for $\Delta M_{F}=\pm1$ transitions. $M_{eg}$ is a transition (electron-spin)
matrix element; $\ B$, is the mean magnetic field ($\approx$ the applied
field) at the sample's surface; and $N$, is the total number of magnetic
vortices in the sample. Most importantly, $S_{2}(\mathbf{q})=(1/N)\left|
\sum_{j}\exp(i\mathbf{q\cdot R}_{j})\right|  ^{2}$ is the vortex lattice
``structure factor.'' We also define a \emph{level-to-level} excitation rate,
$R_{GE}(v,x)$, by summing Eq. \ref{rate coefficient} over (Zeeman) final
states and averaging over initial states:%

\begin{equation}
R_{GE}(v,x)=\frac{1}{3}\sum_{g,e}R_{ge}(v,z) \label{level rate}%
\end{equation}

Here, the upper case letters, $G$, and $E$, refer to the ground and
excited\emph{\ levels}, respectively.

We jump ahead a little and show in Fig. 10 \emph{\ first order perturbation
theory} predictions for the excitation probabilities of atoms passing at
different heights above the 10.7 G vortex lattice discussed previously. These
curves were generated by multiplying the rates of Eq. \ref{level rate} by the
time it takes an atom with the given velocity to pass over the sample. This
particular figure used a structure factor that yields a good fit with our
measurements, though, at this point the details of the calculation are
unimportant and we are using the figure to illustrate a couple of general
features. Specifically (1) the actual signal from the experiment is due to
atoms passing extremely close to the sample's surface --- atoms passing at
heights over 1 $\mu$m are negligibly excited. And (2) the simple perturbation
theory treatment ``predicts''\ excitation probabilities greater than unity for
atoms that pass close to the sample. This indicates that for some heights, the
transitions are saturated and it is necessary to go beyond first order
perturbation theory and include saturation in our description of the
excitation process.

In analyzing the experiment, those atoms passing \emph{extremely} close to the
surface would seem to be problematic for an even deeper reason. Very close to
the sample's surface, the magnetic field varies widely and treating the
fluctuating field due to the vortices as a perturbation on top of the nominal
bias field at the surface is not justified. Fortunately, these atoms do not
contribute to the signal, but rather are pulled into the sample by van der
Waals forces. We include the van der Waals force in our modeling by using it
to provide a lower cutoff \ to heights above the sample included in predicting
signals. This cutoff is on the order of 0.2 $\mu$m and depends on the atoms'
velocity. Still, referring to Fig. 10 even with a lower cutoff of 0.2 $\mu$m ,
saturation effects are seen to be important.

We use a Master Equation approach to include saturation in our model. As an
atom crosses the sample, the probability of its excitation by the vortex
lattice to a state $e$ in the $F=2$ manifold is assumed to satisfy the
differential equation (Master Equation):%

\begin{equation}
\frac{dP_{e}(v,z,t)}{dt}=\sum_{g^{\prime}}R_{g^{\prime}e}(v,z)\left[
P_{g^{\prime}}(v,z,t)-P_{e}(v,z,t)\right]  \label{Master Equation}%
\end{equation}
Where the sum is over the $F=1$ states, $R_{g^{\prime}e}(v,z)$ is given by Eq.
\ref{rate coefficient} and we have used the fact that the state-to-state
\emph{excitation} rate is equal to the corresponding \emph{deexcitation} rate.
Corresponding equations describe the occupation probability of the lower,
$F=1$ states. The excitation probability for an atom is readily found by
integrating these (eight) coupled equations for the time the atom is over the
sample with the condition that when the atom initially encounters the sample,
$P_{g}(v,z,t=0)=1/3$ for each of the three $F=1$ states. For our experimental
conditions, intra-level transitions --- e.g. transitions between states in the
ground level --- should be weak.

Cohen-Tannoudji et al.\cite{cohentannoudgi} discuss the justification of the
Master Equation approach in describing radiative processes (Einstein A and B
coefficients) and much of that discussion is readily adaptable to the present
case. In particular, a ``coarse graining'' of the excitation process allows
using a rate coefficient to describe coherent excitation. The basic idea is
that the net excitation consists of \ sum coherent excitations that are,
individually, independent of one another \ --- they add incoherently. To carry
out ``coarse graining'' in the vortex-lattice-excitation-of-atoms problem ,
the coherence time, $\tau_{coh}$ , of the vortex-lattice field that excites
the atom needs to be much shorter than, $\tau_{ex}$, the time it takes that
field to coherently excite the atom. In other words, an atom should be only
weakly excited ($\Delta P_{e}\ll1$) during the time it travels over the sample
a distance equal to the translational correlation length of the \ vortex
lattice. For even the most strongly excited atoms that contribute to the
signal (i.e. those at the lower cutoff height), this condition is met.

Additionally, using Eq. \ref{rate coefficient} for transition rates in the
Master Equation(s) \emph{implicitly\ }assumes that the vortex lattice is
homogenous across the entire sample. For example, if the sample's vortex
lattice \emph{actually} consisted of two large domains with very different
(local) structure factors, our treatment would need to be extended.
Specifically, we assume that the vortex driven excitation \emph{rate},
$R_{g^{\prime}e}(v,z)$ (coarse-grained --- averaged over distances large
compared to the vortex lattice translational correlation length) is uniform
across the sample. In our work, the most likely violation of this condition
results from edge effects. We experimentally investigated this issue by making
magneto-optic images of the flux density of the sample and we know from micro
hall array measurements, that, for the conditions of Fig. 5, the flux density
at the surface is uniform to within a couple of per cent across for the
central 2/3's of the sample (it is slightly lower near the sample edges). Of
course, even uniformity of the field across the sample, does
not\emph{\ necessarily} mean the vortex lattice itself is homogenous.

For the results presented in this paper, we treat the Zeeman states within the
hyperfine levels individually to first order. This is important since the
Zeeman shift of the hyperfine states broadens as well as shifts the 'average'
$F=1\rightarrow F=2$ transition. In order to account for saturation, we
generalize Eq. \ref{Master Equation} to the extent possible and work with a
single, level-to-level Master equation,%

\begin{equation}
\frac{dP_{E}(v,z,t)}{dt}=R_{GE}(v,z)\left[  P_{G}(v,z,t)-\frac{3}{5}%
P_{E}(v,z,t)\right]  =R_{GE}(v,z)\left[  1-\frac{8}{5}P_{E}(v,z,t)\right]  .
\label{level rate expression}%
\end{equation}
This two-level Master equation is the principal result of the section. Here,
$R_{GE}(v,z)$ is the level-to-level transition rate, Eq. \ref{level rate},
$P_{E}$ and $P_{G}$, are the occupation probabilities of the\emph{\ levels
}(summed over the Zeeman states) and the $\frac{3}{5}$ factor insures that for
strong saturation, the levels will be statistically populated according to
their degeneracies. Note: while Eq. \ref{level rate expression} \emph{\ }%
follows from Eqs. \ref{level rate} and \ref{rate coefficient} in the
unsaturated and strongly-saturated limits, for the general case, this rate
equation for the levels \emph{cannot }be\emph{\ rigorously} derived from the
system of rate equations linking the individual states. However, we have made
several comparisons between this model and the full solution of the
Schr\"{o}dinger Eq. and generally find agreement to much better than 10\%.
Fig. 11 shows one such comparison.\ Importantly, the model results require
less than 1/1000th the computer time of the Schr\"{o}dinger equation
solutions. We conclude that, for the purpose of comparing with our
measurements, the two-level Master equation model based on Eq. \ref{level rate
expression} provides an adequate description of the excitation process.

\subsection{Modeling the structure factor}

We showed in the previous section that the atomic beam signal is intimately
related to the two-dimensional structure factor, $S_{2}(\mathbf{q}%
)=(1/N)\left|  \sum_{l=1}^{N}e^{i\mathbf{q\cdot R}_{l}}\right|  ^{2}$, for an
array of $N$ vortices at positions $\left\{  \mathbf{R}_{l}\right\}  $. The
purpose of this section is to develop a general framework for describing the
structure factor for a wide range of potential vortex lattices. For a
homogeneous system, $\ S_{2}(\mathbf{q})$ can be expressed as \cite{Ziman79}%

\begin{equation}
S_{2}(\mathbf{q})=1+\int d^{2}re^{i\mathbf{q\cdot r}}f(\mathbf{r)},
\end{equation}
where $f(\mathbf{r})$ is the probability distribution for finding a pair of
vortices separated by a distance $\mathbf{r}$.\ The normalized pair
distribution function is $g(\mathbf{r})=f(\mathbf{r})/n$, where $n=B/\phi_{0}$
is the average number density of vortices. In the following, we model the
distribution function $f(\mathbf{r})$ for a lattice spanned by the primitive
vectors $\mathbf{r}_{1}$ and $\mathbf{r}_{2}$, and assume that vortex
displacements relative to perfect crystalline order is described by a Gaussian
distribution. See Fig. 12. We write%

\begin{equation}
f(\mathbf{r})=\sum\nolimits_{l,m}^{^{\prime}}\frac{1}{2\pi\sigma(r_{lm})^{2}%
}\exp\left(  -\frac{(\mathbf{r-}l\mathbf{r}_{1}\mathbf{-}m\mathbf{r}%
_{2}\mathbf{)}^{2}}{2\sigma(r_{lm})^{2}}\right)  ,
\end{equation}
where $r_{lm}=\left|  l\mathbf{r}_{1}+m\mathbf{r}_{2}\right|  $ and the sum
runs over $l$ and $m$ except $l=m=0$. The Fourier transform of $f(r)$ is
evaluated and the structure factor is
\begin{equation}
S_{2}(\mathbf{q})=1+\sum\nolimits_{l,m}^{^{\prime}}e^{-i\mathbf{q\cdot
(}l\mathbf{r}_{1}\mathbf{+}m\mathbf{r}_{2}\mathbf{)}}e^{\mathbf{-}q^{2}%
\sigma(r_{lm})^{2}/2}. \label{GaussianS2}%
\end{equation}
Here the functional form of the displacement $\sigma(r)$ is used to
parametrize the range and magnitude of correlations in the vortex array. In
this work we have used a displacement of the form
\begin{equation}
\sigma(r)=\sigma_{0}(r/a_{0})^{p}, \label{displacementfunction}%
\end{equation}
where $a_{0}=\sqrt{\phi_{0}/B}$ is the average vortex separation. For the
numerical evaluation of $S_{2}(q)$ from expression \ref{GaussianS2}, one has
to sum over a sufficiently large lattice in order to ensure convergence.

So far we have only included positional disorder in the expression for the
structure factor. For the subsequent analysis, we also include orientational
disorder by averaging the structure factor over different orientations of the
unit cell spanned by $\mathbf{r}_{1}$ and $\mathbf{r}_{2}.$ In practice this
is done by computing $S_{2}(q)$ for a particular $\mathbf{r}_{1}$ and
$\mathbf{r}_{2}$ and then averaging over orientations with a Gaussian
distribution with a standard deviation $\Delta\theta$ around $0^{\circ}$, i.e.
the orientationally averaged structure factor is
\begin{equation}
\overline{S}_{2}(\mathbf{q})=\frac{1}{\sqrt{2\pi}\Delta\theta}\int_{-\infty
}^{+\infty}d\theta S_{2}\left(  \mathbf{q\prime}(\theta)\right)  \exp\left(
-\frac{\theta^{2}}{2(\Delta\theta)^{2}}\right)  , \label{averagedS2}%
\end{equation}
where $\mathbf{q}^{\prime}(\theta)$ is the wave vector $\mathbf{q}$ rotated an
angle $\theta$.

The model for $S_{2}(\mathbf{q})$ $\ $as outlined above in expressions
(\ref{GaussianS2}) --- (\ref{averagedS2}) provides a quite general framework
for describing different classes of \ disorder, \ ranging from crystalline
long range order ($p=0,\Delta\theta=0)$, via powder correlations
($p=0,\Delta\theta\rightarrow\infty)$ and hexatic order ($p=0.5,\Delta
\theta\ll\pi/6)$ to isotropic liquid-like order $(p=0.5,\Delta\theta
\rightarrow\infty)$. The main fitting parameters are the nearest neighbor
displacement $\sigma_{0}$, the exponent $p$ and the orientational disorder
$\Delta\theta$. The penetration depth is sometimes taken as a free parameter
and other times fixed using values in the literature.

\section{Formal analysis of vortex correlations in YBCO: a case study}

Next, we use the modeling framework for a quantitative analysis of the vortex
lattice correlation in a high quality YBCO single crystal. Comparison between
model predictions and experimental data enables us to extract information on
the overall symmetry and orientation of the vortex array, the range and
magnitude of vortex correlations, and the London penetration depth,
$\lambda_{\text{ab}}$.

We reconsider the data shown in Fig. 5. Formally, we \emph{assume }that the
vortex lattice is accurately described by a single, sample-wide
autocorrelation function and does \emph{not}, for example, consist of two
different domains. Note: this also ignores possible edge effects. \ We further
\emph{assume }that the underlying vortex lattice in the sample, ignoring
disorder, can be described by two primitive lattice vectors, $\mathbf{r}_{1}$
and $\mathbf{r}_{2}\mathbf{,}$ as discussed above. This is a much weaker
assumption than in the Overview (Section 2) where we restricted consideration
to triangular lattice variants. For example, if the vortex lattice
\emph{actually} had square or rectangular symmetry, we would discover this
fact in the course of the present analysis.

A fit to the data proceeds at two levels: i) The overall symmetry and
orientation of the vortex lattice unit cell (Fig.13) is determined by the two
length scales of the ABMR peaks \emph{and} their relative intensities. ii) The
shape and strength of the ABMR signal contain information on the magnitude and
range of the vortex correlation function, as well as the penetration depth
$\lambda_{\text{ab}}.$ We retrieve that information by comparing model
predictions for various putative correlation functions with the experimental
data. Note that the vortex array correlation \emph{cannot} be extracted by
simple curve-fitting to the diffraction peaks, in contrast to neutron
diffraction data, since the peaks are \emph{broadened} due to saturation of
the ABMR transitions (section 4.2). For the same reason, there is no simple
analytic relation between the penetration depth and the total signal strength.

We consider the oblique unit cell specified by the magnitudes of the primitive
lattice vectors, $\mathbf{r}_{1}$ and $\mathbf{r}_{2}$, the angle $\theta$
that $\mathbf{r}_{1}$ makes with the atomic beam direction (x-axis) and the
angle $\beta$ between $\mathbf{r}_{1}$ and $\mathbf{r}_{2}$. These four
parameters have to satisfy the following criteria: a) The two shortest
projected lattice row spacings, $d_{1}$and $d_{2}$ in Fig. 13, have to equal
the ABMR peak positions 1.24 $\mu$ m and 1.68 $\mu$ m. b) The unit cell area
be correct, $r_{1}r_{2}\sin\beta=\phi_{0}/B$ .\ And c) The unit cell has to
reproduce the relative intensities of the two ABMR peaks. This last criterion
is less ``absolute'' than a) and b) in the sense that the interdependencies it
imposes on the lattice parameters depends, weakly, on the chosen model of
disorder. The best-fit-model vs. data comparison is shown in Fig. 13. The unit
cell is described by $r_{1}=1.4\mu$ m and $r_{2}=1.6\mu$ m at an angle
$\beta=62^{\circ}$. The shortest primitive lattice vector, $\mathbf{r}_{1}$,
is rotated an angle $\theta=18^{\circ}$ with respect to the crystalline a-axis
of the sample. These results are consistent with the less formal discussion in
the overview of section 2.

A fit to the shape and overall strength of the ABMR signal uses the modeling
framework developed above. For a given set of lattice parameters, we compute
the structure factor $S_{2}(\mathbf{q})$ (Eqs. \ref{GaussianS2} --
\ref{averagedS2}) parametrized by the displacement relative to perfect order
$\sigma(r)=\sigma_{0}(r/a_{0})^{p}$, and the angular disorder $\Delta\theta$.
The structure factor $S_{2}(\mathbf{q})$ is used together with the penetration
depth $\lambda_{\text{ab}}$ as an input for computing the hyperfine excitation
rate coefficients, Eqs. \ref{rate coefficient}, \ref{level rate}and the
transition probability, Eq.\ref{level rate expression} as a function of
velocity $v$ for a given height $z$. Finally, the transition probability is
averaged over the relevant range of heights. Model predictions are generated
for a range of values of the input parameters. The best fit model shown in
Fig. 13 is for $p=0.5\pm0.1$, $\sigma_{0}=\left(  0.14\pm0.01\right)  a_{0}$,
and angular disorder $\Delta\theta\leq2^{\circ}.$

Fig. 13 shows that the overall strength of the signal at 10 K is best matched
for a penetration depth, $\lambda_{\text{ab}}=140\pm20$ nm. This is consistent
with typical values for the low temperature penetration depth found by other
workers. By combining this result with the the distortion factor found in
section 2, our measurements independently provide estimates of the individual
penetration depths, $\lambda_{a}=151\pm20$ nm and $\lambda_{b}=130$ $\pm20$
nm. These values are virtually identical to those deduced from Johnson et
al's\cite{JFL99} small angle neutron scattering work, $\lambda_{a}=150$ nm and
$\lambda_{b}=127$ nm, though given the error bars of our measurements, the
level of agreement is somewhat fortuitous. Similarly, our values are in
harmony with the polarized light scattering measurements of Wang et
al,\cite{WTR98} $\lambda_{a}=160$ nm, $\lambda_{b}=117$ nm and appear
consistent with the Josephson-tunneling results of Sun et al,\cite{SHK95} who
report values for $\lambda_{a}=161$ to $270$ nm and $\lambda_{b}=90$ to$\ 174$
nm for several different samples.

\section{Concluding Remarks}

We conclude with some general observations about the ABMR technique, itself.

The strengths and weaknesses of the ABMR are largely complementary to those of
other imaging-type vortex lattice diagnostic tools. Weaknesses include, in
contrast to neutron diffraction measurements, the technique directly provides
only 1-D information, it probes vortices only at a superconductor's surface
and as currently implemented it is restricted to relatively low fields. In
contrast with decoration experiments or electron holography, the atomic beam
technique provides sample averaged information about the vortex lattice: it
cannot look at individual vortices or even domains within the vortex lattice.

On the positive side, some cases --- e.g. isotropic lattices or lattices with
sample-wide order --- the 1-D measurement can lead to a 2-D description of the
vortex lattice. The atomic beam technique does \emph{not} require especially
thin or especially smooth samples. Nor does it require a large quantity of
sample material --- c.f. neutron diffraction studies. In contrast to the
``one-shot'' nature of decoration experiments and the heroic integrations
sometimes required for neutron diffraction measurements, atomic beam
measurements are made in near-real time and can follow the evolution of a
vortex lattice on a time scale of minutes. Measurements are possible in the
presence of transport currents and are possible even if vortices are moving.
Finally, to date, it is the \emph{only} imaging-type diagnostic tool that has
had sufficient sensitivity to study vortex lattices near T$_{c}$\cite{nature}.

In many ways, atomic beam technique is similar to neutron diffraction. Both
provide information on the vortex lattice in $k-$ space. Both provide
sample-averaged information. The discussion leading up to Eq. \ref{rate
coefficient} is very similar to that needed to predict and analyze signals
from neutron diffraction experiments. \cite{ForganPaulMook90} Signal strengths
for both types of experiments (to first order) $\sim$ $\lambda_{ab}^{-4}$.
Approaching $T_{c}$ from below, the penetration depth diverges and for both
neutron diffraction and AMBR experiments this leads to sharply decreasing
signals and increasingly difficult measurements. Why is it that the atomic
beam method is able to make measurements near $T_{c}$? The sensitivity of the
atomic beam method relative to neutron diffraction is a consequence of the
fact that in the atomic beam method, the signal is produced by the interaction
of the Bohr magneton of an atom's valence electron with the vortex lattice
field. In neutron diffraction, \ it is the much smaller nuclear magneton--
magnetic field interaction that generates the signal.

We gratefully acknowledge the technical assistance of John Spaulding, the OSU
model shop, and Jeff Fox and useful discussions with Tom Lemberger. We thank
Mike McElfresh/Eli Zeldov for providing the Hall array. This work was
supported by the PYI program of NSF and the MISCON program of DOE. HH
acknowledges support from the University of Oslo and the Norwegian Research Council.

\section{\bigskip Appendix: Excitation rate coefficient for an atom over a
vortex lattice}

For our models we need to find the sample-averaged excitation rate
$F=1\rightarrow F=2$ of an atom with a given velocity and height traveling
above vortex lattice. We begin with the first order perturbation theory
calculation of excitation out of the $F=1$ level. We take as a basis the
hyperfine (variables=electronic spin, nuclear spin) energy eigenstates for the
mean field \ that we measure at the sample's surface. The quantization axis is
given by this field --- i.e. is perpendicular to the sample surface. The
magnetic resonance transition is a magnetic dipole transition with selection
rules $\Delta M_{F}=\pm1,0$ and thus within first order perturbation theory,
the problem of excitation by the vortex lattice of the initial, $F=1$ level,
reduces to nine uncoupled two-state problems corresponding to the nine allowed
$F=1\rightarrow F=2$ transitions. We consider an individual transition between
a specific, initial magnetic state, $\left|  g\right\rangle ,$ in the lower
level and a specific final magnetic state, $\left|  e\right\rangle $, in the
upper level. The interaction Hamiltonian has a non-zero matrix element,
$H_{eg}^{\prime}(t)$ , for only one term (at most) of the sum in Eq.
\ref{interaction hamiltonian} and will driven by only one spherical component
of the field. The transition probability that an atom, initially in $\left|
g\right\rangle $ is excited to state $\left|  e\right\rangle $ is to first
order in the interaction:
\begin{align}
P_{ge}^{(1)}  &  =\frac{1}{\hbar^{2}}\left|  \int_{-\infty}^{+\infty}%
H_{eg}^{\prime}(t)e^{i\omega_{eg}t}dt\right|  ^{2}\label{tdpt}\\
&  =4\left(  \frac{\mu_{B}}{\hbar}\right)  ^{2}|M_{eg}|^{2}\left|
\int_{-\infty}^{+\infty}B_{-j}(t)e^{i\omega_{eg}t}dt\right|  ^{2}.\\
&  =4\left(  \frac{\mu_{B}}{\hbar}\right)  ^{2}|M_{eg}|^{2}\left|
\int_{-\infty}^{+\infty}B_{j}(t)e^{-i\omega_{eg}t}dt\right|  ^{2}%
\end{align}
Here $H_{eg}^{\prime}(t)$ is the matrix element of the interaction
Hamiltonian, $M_{eg}=\left\langle e\left|  S_{j}\right|  g\right\rangle $ is
the non-zero electron spin operator matrix element and $\omega_{eg}$ is the
magnetic resonance transition's frequency, $=2\pi\cdot462MHz$ + the
transition's Zeeman shift for the average field near the sample's surface. The
final result, here, reflects the fact that the excitation probability is
proportional to the power spectrum of the field seen by the atom at the
resonant frequency of the transition.

For an atom travelling in the $x$-direction with velocity $v$, at a height $z
$ above the sample's surface and a transverse displacement $y$ from the center
of the sample, the\emph{\ temporal} Fourier transform $B_{j}(\omega)$ of the
magnetic field $B_{j}(t)$ in the moving reference frame of the atom is related
to the \emph{spatial} Fourier transform $B_{j}(q_{x},y,z)$ of the magnetic
field $B_{j}(x)$ along the atomic trajectory in the lab frame:
\begin{equation}
\int_{-\infty}^{+\infty}B_{j}(t)e^{-i\omega_{eg}t}dt=\frac{1}{v}\int_{-\infty
}^{+\infty}B_{j}(x,y,z)e^{-iq_{x}x}dx=\frac{1}{v}B_{j}(q_{x},y,z),
\end{equation}
where $q_{x}=\omega_{eg}/v.$ The distance scale probed by atoms with velocity,
$v$, is $l=2\pi/q_{x}=v\cdot2\pi/\omega_{eg}$. The first order transition
probability can now be written:
\begin{equation}
P_{eg}^{(1)}=\frac{4}{v^{2}}\left(  \frac{\mu_{B}}{\hbar}\right)  ^{2}\left|
M_{eg}\right|  ^{2}\left|  B_{j}(q_{x},y,z)\right|  ^{2}. \label{pert}%
\end{equation}
Here $B_{j}(q_{x},y,z)$ is understood as the partial Fourier transform of
$B_{j}(x,y,z)$ for a particular trajectory given by $y$ and $z$. In the actual
experiment, we measure the transition probability averaged across a beam's
transverse dimensions, $y$ and $z$. $\left|  B_{j}(q_{x},y,z)\right|  ^{2}$
can be expressed in terms of the 2D Fourier transform $B_{j}(\mathbf{q},z)$ of
the previous section:
\begin{equation}
\left|  B_{j}(q_{x},y,z)\right|  ^{2}=\left|  \frac{1}{2\pi}\int_{-\infty
}^{+\infty}dq_{y}B_{j}(\mathbf{q},z)e^{iq_{y}y}\right|  ^{2}.
\end{equation}
The average of this quantity across the width if the sample is
\begin{equation}
\frac{1}{L_{y}}\int dy\left|  B_{j}(q_{x},y,z)\right|  ^{2}=\left(
\frac{1}{2\pi}\right)  ^{2}\frac{1}{L_{y}}\int\limits_{-L_{y}/2}^{L_{y}%
/2}dy\int\limits_{-\infty}^{+\infty}dq_{y}\int\limits_{-\infty}^{+\infty
}dq_{y}^{,}B_{j}(\mathbf{q},z)B_{j}^{\ast}(\mathbf{q}^{,},z)e^{i(q_{y}%
-q_{y}^{^{\prime}})y}.
\end{equation}
Assuming that $L_{y}$ is large we use $\int dy\,e^{i(q_{y}-q_{y}^{^{\prime}%
})y}=2\pi\delta(q_{y}-q_{y}^{^{\prime}})$and obtain the average transition
probability for atoms travelling at height $z,$
\begin{equation}
P_{ge}^{(1)}(q_{x},z)=\frac{2}{\pi v^{2}L_{y}}\left|  M_{eg}\right|
^{2}\left(  \frac{\mu_{B}}{\hbar}\right)  ^{2}\int_{-\infty}^{+\infty}%
dq_{y}\left|  B_{j}(\mathbf{q},z)\right|  ^{2}. \label{1storderaverage}%
\end{equation}
Including the expression for the magnetic field structure near the sample
surface (\ref{fieldstructure}), we find \/the average first order transition
probability for atoms travelling near a vortex array with speed $v$ along the
$x$-axis at height $z$.
\begin{equation}
P_{ge}^{(1)}(v,z)=\eta_{j}\frac{4\left|  M_{eg}\right|  ^{2}}{v^{2}}\left(
\frac{\mu_{B}}{\hbar}\right)  ^{2}\frac{\phi_{0}^{2}}{2\pi\lambda_{\text{ ab}%
}^{4}}\frac{1}{L_{y}}\int\limits_{-\infty}^{+\infty}dq_{y}\frac{e^{-2qz}%
}{(\alpha(\alpha+q))^{2}}\left|  \sum\limits_{l=1}^{N}e^{i\mathbf{q\cdot
R}_{l}}\right|  ^{2}, \label{1storderaverage2}%
\end{equation}
Where $\eta_{j}=1$ for $\Delta M_{J}=0$ transitions in Eq. \ref{interaction
hamiltonian} and $\eta_{j}=\frac{1}{2}$ for $\Delta M_{J}=\pm1$ transitions.
Rewriting this in terms of the two-dimensional vortex array structure factor,
$S_{2}(\mathbf{q})=(1/N)\left|  \sum_{j}\exp(i\mathbf{q\cdot R}_{j})\right|
^{2}$ (note: the structure factor is, to within a constant, the Fourier
transform of the vortex lattice autocorrelation function)$:$
\begin{equation}
P_{ge}^{(1)}(v,z)=2\eta_{j}\left|  M_{eg}\right|  ^{2}\frac{L_{x}}{\pi v^{2}%
}\frac{B\phi_{0}}{\lambda_{\text{ab}}^{4}}\left(  \frac{\mu_{B}}{\hbar
}\right)  ^{2}\int\limits_{-\infty}^{+\infty}dq_{y}\frac{e^{-2qz}}%
{(\alpha(\alpha+q))^{2}}S_{2}(\mathbf{q}) \label{1storder}%
\end{equation}
Here we have used $N\phi_{0}/(L_{x}L_{y})=B$, where $B$ is the flux density,
$L_{x}L_{y}$ is the sample area and $N$ is the number of vortices. To find the
\emph{total }excitation by the vortex lattice this result should be averaged
over initial states and summed over final state.

Finally, we identify the average \emph{rate} for the excitation process
$g\rightarrow e$. This is given by the transition probability (\ref{1storder})
divided by the time $\tau=L_{x}/v$ it takes an atom moving with velocity, $v$,
to pass over the sample.
\begin{equation}
R_{ge}(v,z)=2\eta_{j}\left|  M_{eg}\right|  ^{2}\frac{1}{\pi v}\frac{B\phi
_{0}}{\lambda_{\text{ab}}^{4}}\left(  \frac{\mu_{B}}{\hbar}\right)  ^{2}%
\int\limits_{-\infty}^{+\infty}dq_{y}\frac{e^{-2qz}}{(\alpha(\alpha+q))^{2}%
}S_{2}(\mathbf{q}), \label{rate expression}%
\end{equation}

\bigskip\newpage

\bigskip

{\large Figure Captions}

\bigskip

FIG. 1. Principle behind Atomic Beam Magnetic Resonance. An atom travelling
through the \emph{spatially inhomogeneous} magnetic field above the surface of
a type II superconductor in the mixed state will experience a
\emph{time-dependent} magnetic field. The frequency spectrum of that field
depends on the spatial pattern of vortices and the velocity of the atom. If
the magnetic field has a frequency component coincident with a magnetic dipole
transition of the atom, that transition may be strongly driven. For a given
transition, atoms travelling at different velocities will be sensitive to
different \emph{spatial} Fourier components of the inhomogeneous magnetic
field. Therefore by measuring the transition probability for a particular
transition \emph{as a function of the atomic velocity}, we can study the
spatial characteristics of the magnetic field associated with the flux line
array and hence the structure of the vortex lattice itself.

\bigskip

FIG. 2. Zeeman manifolds for the relevant energy levels of $^{39}$K. Shown are
manifolds for the three lowest electronic states. Faint horizontal lines show
nearby hyperfine states that are not \emph{directly} used in the experiment.
The energy scale is extremely distorted. Dashed lines indicate transitions
that result from spontaneous emissions. The detection transition indicated is
the cycling transition that is responsible for most of the fluorescence
signal. The magnetic resonance (MR) transition that is driven by the vortex
lattice, as discussed in the text, actually consists of nine allowed
state-to-state excitations between the $F=1$ and $F=2$ levels. Similarly, the
figure shows only one of the several state-to-state transitions transitions
that are driven by the pumping laser.

\bigskip

FIG. 3. Experimental layout. The apparatus consist of three functionally
distinct sections: (1) State preparation, where a thermal beam \ ($T\simeq
400^{\circ}$ C)of potassium atoms is created and optically pumped into the
$F=1$ level; (2) A cryogenic sample region, where the atomic beam passes close
to the surface of the superconducting sample in the mixed state and magnetic
resonance transitions may be driven. For most of this work, the sample
consists of a thin platelet of YBCO in a $\sim10$ G field perpendicular to its
surface (crystalline $c-$ axis). (3) A detection region in which atoms that
have been excited to the $F=2$ manifold are detected as a function of their
velocity. Excited atoms are detected \emph{via} laser induced fluorescence.
The Doppler shift of the laser driven transition is used to discriminate atom
velocities. Small signals and large backgrounds require an involved detection
scheme, as described in the text.

\bigskip

FIG. 4. ABMR data acquired for different values of the bias field. Vortex
lattices are created by ``field cooling.'' I.e. the sample is cooled through
its transition temperature in the presence of the indicated bias magnetic
field. These data were taken for $T=$10 K. They show the expected trend: for
stronger fields, vortices are closer together.

\bigskip

FIG. 5. Data for a sample in a field of 10.7 G. These data were acquired for a
sample temperature of 10 K. For this case, the 10.7 G refers to the field
actually measured by a miniature Hall probe array at the surface of the
sample. The two peaks --- at 1.24 $\mu$ and 1.68 $\mu$ --- describe a
triangular vortex lattice with sample wide order whose symmetry axes do
\emph{not} lie along the direction of the atomic beam.

\bigskip

FIG. 6. Triangular vortex lattices. The strongest peaks in ABMR spectra are
due to the coherent excitation of atoms by rows of nearest neighbor vortices.
For triangular lattices, there are three sets of such rows as shown on the
figures. For each set of rows, a bar indicates the distance atoms travel
between rows; these distances correspond to peaks in the spectra. Fig A) shows
the case for the atomic beam traveling along a symmetry axis of the vortex
lattice: this would create one sharp peak. Fig B) shows a vortex lattice that
is tilted with respect to the atomic beam. Specifically, Fig B) shows a tilt,
$\theta=15^{\circ}$ --- the orientation of the vortex lattice measured in FIG.
5. The dashed [dotted] rows produce the $d_{1}=$1.68 $\mu$ [$d_{2}=$1.24 $\mu
$] peaks in that figure. At the top, a third peak is predicted but its
distance is larger than the range of sensitivity of the present experiment.

\bigskip

FIG 7. Niobium data from ``I.'' The broad asymmetric peak is characteristic of
a vortex lattice made up of randomly oriented domains within which there is
short range translational order. The smooth curve is a model prediction for
such a vortex lattice.

\bigskip

FIG. 8. A gallery of ABMR signals predicted for several forms of vortex
matter. (A) Gas of uncorrelated vortices, (B) liquid with short range
translational and orientational order, (C) Powder correlations due to randomly
oriented vortex crystallites (D) Hexatic vortex glass with sample wide
orientational order and short range translational order, oriented with a
primitive lattice vector along the atomic beam. (E) is (D) rotated by
15$^{\circ}.$ (F) Excitation due to a \ triangular near-crystalline vortex
array with a primitive lattice vector oriented along the atomic beam.

\bigskip

FIG. 9. Representative raw experimental signals and reduced data.
Experimentally, the top curves are measurements of laser induced fluorescence.
The ``pumped cold'' data contain the signal and the ``pumped warm'' data are
the background. The ``unpumped distribution'' is used to normalize the data
and shows the sensitivity of our apparatus to different velocity classes of
atoms. The velocity scale on the horizontal axis is determined from the
Doppler shift of the detection laser and may be immediately converted to the
distance scale of the other figures using $d=v/f.$ By reducing the data, as
discussed in the text, we directly measure the absolute excitation probability
of the atoms by the vortex lattice.

\bigskip

FIG. 10. Calculated first order excitation ``probability'' for atoms passing
at different heights above a vortex lattice. The calculation is from the
vortex lattice model that best reproduces the experimental data of FIG. 5.
Note that essentially all of the signal comes from atoms passing within 1
$\mu$ of the surface. The importance of including saturation effects in the
modeling is seen where the first order theory predicts probabilities
$>$
1 for atoms passing close to the surface

\bigskip

FIG. 11. Testing the two-level Master equation model by comparing it with an
exact solution of the Schr\"{o}dinger equation calculation. Calculations are
done for 690 m/s atoms passing 0.2 $\mu$ above a triangular lattice. The model
lattice was disordered by adding to each vortex a random displacement,
averaging 0.1 lattice spacings, from its nominal\ triangular lattice position.
The jagged curve is a simulation in which the exact time dependent
Schr\"{o}dinger equation of the ground state hyperfine manifold is solved for
200 trajectories over the model vortex lattice. The smooth curve is the time
integration of Eq. \ref{level rate expression} --- the model used in this
work. By construction, the two-level Master equation model is \emph{expected}
to agree with the exact calculation in the perturbative and the fully
saturated limits. Empirically in this and other comparisons, we find good
agreement at \emph{all} times. The exact solution took over 1000 times longer
to run than the Master equation calculation.\bigskip

FIG.12. Contour plot of a model vortex pair distribution function. Deviations
from perfect crystalline order is described by a Gaussian distribution with a
separation dependent variance $\sigma(r)=\sigma_{0}(r/a_{0})^{p}$, where
$a_{0}=\sqrt{\phi_{0}/B}$ is the average vortex spacing. The pair distribution
shown is for $\sigma_{0}=0.14$ and $p=0.5.$ The unit cell is specified by the
magnitudes of the primitive lattice vectors $\mathbf{r}_{1}$ and
$\mathbf{r}_{2}$, the angle $\theta$ that $\mathbf{r}_{1}$ makes with the
atomic beam direction (x-axis) and the angle $\beta$ between $\mathbf{r}_{1}$
and $\mathbf{r}_{2}$.

FIG 13. Best-fit model overlaying the data of FIG. 5. The best fit model
lineshape (dark smooth line) is for a correlation function with $\sigma
_{lm}=\sigma_{0}x_{lm}^{p}$ where $p=0.5$ and $\sigma_{0}=0.14a_{0}$. The
overall signal strength is best fit assuming a penetration depth $\lambda
_{ab}=140$ nm. The upper curve and lower lighter curves show model predictions
for penetration depths of $100$ nm and $180$ nm, respectively.

\bigskip
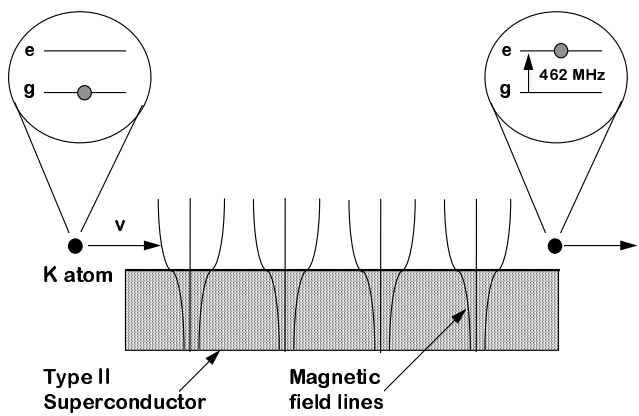
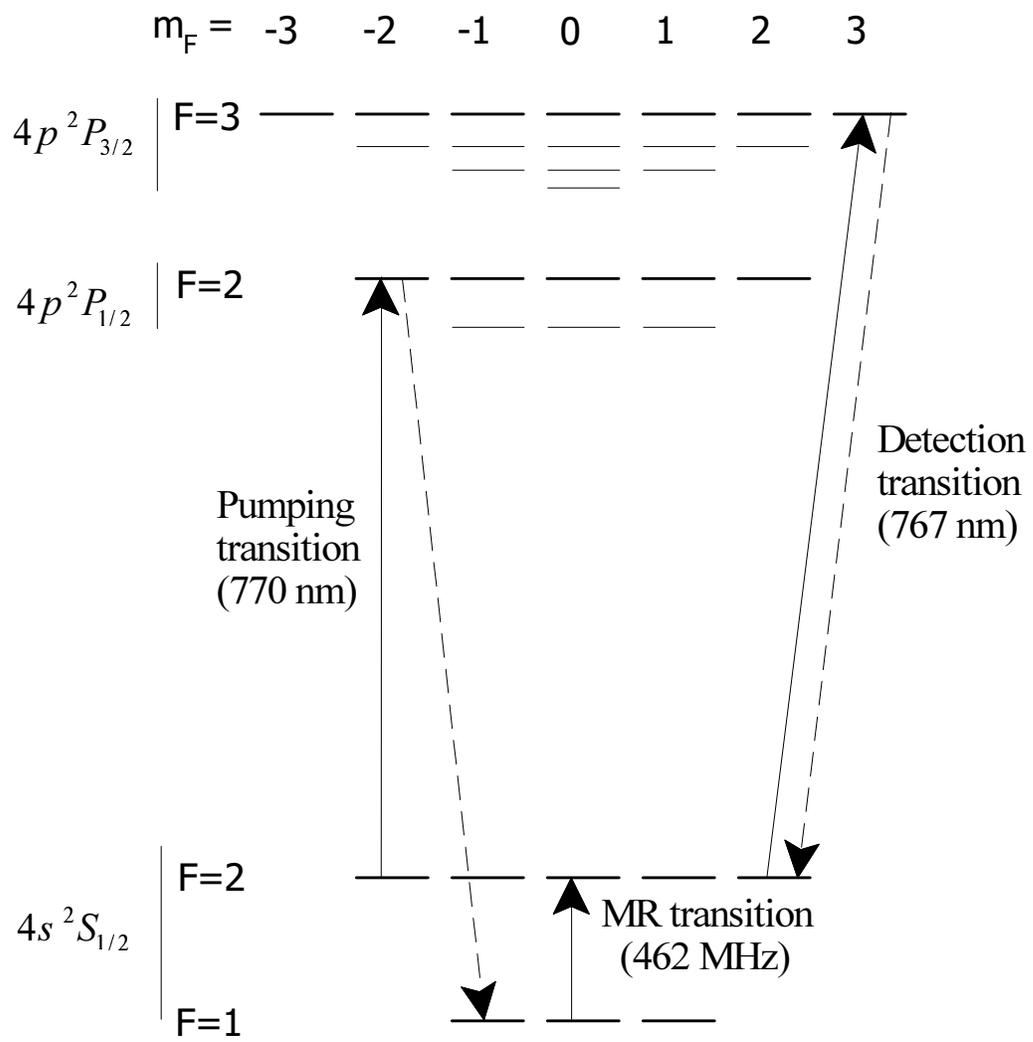
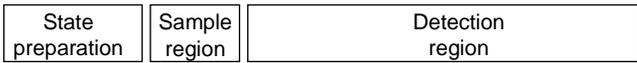
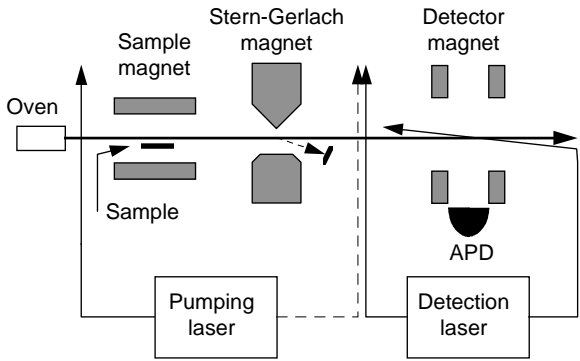
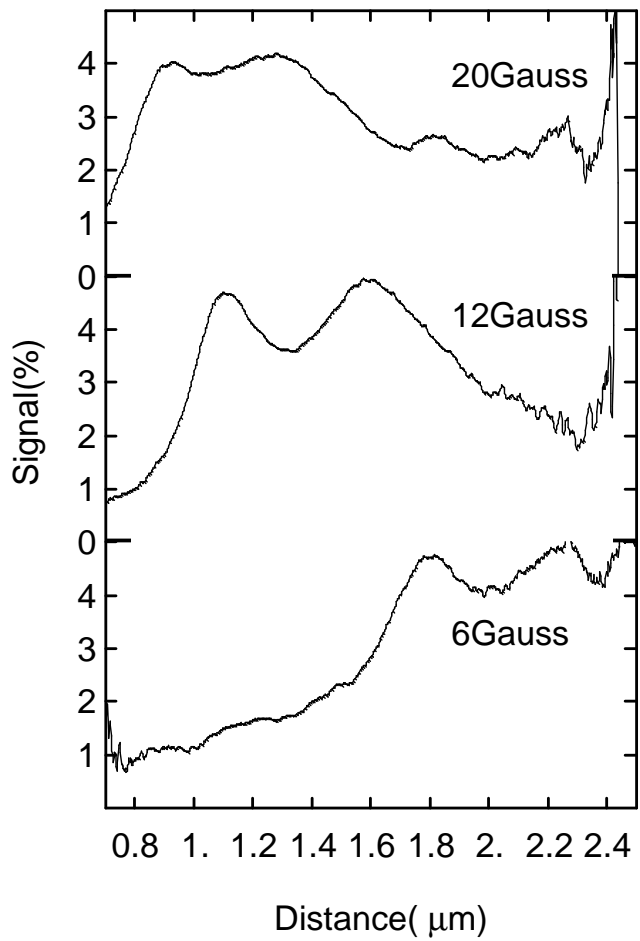
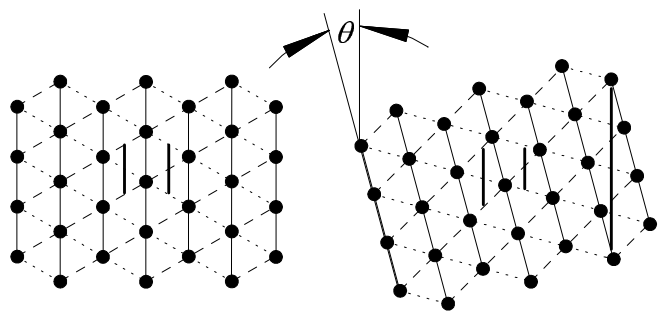
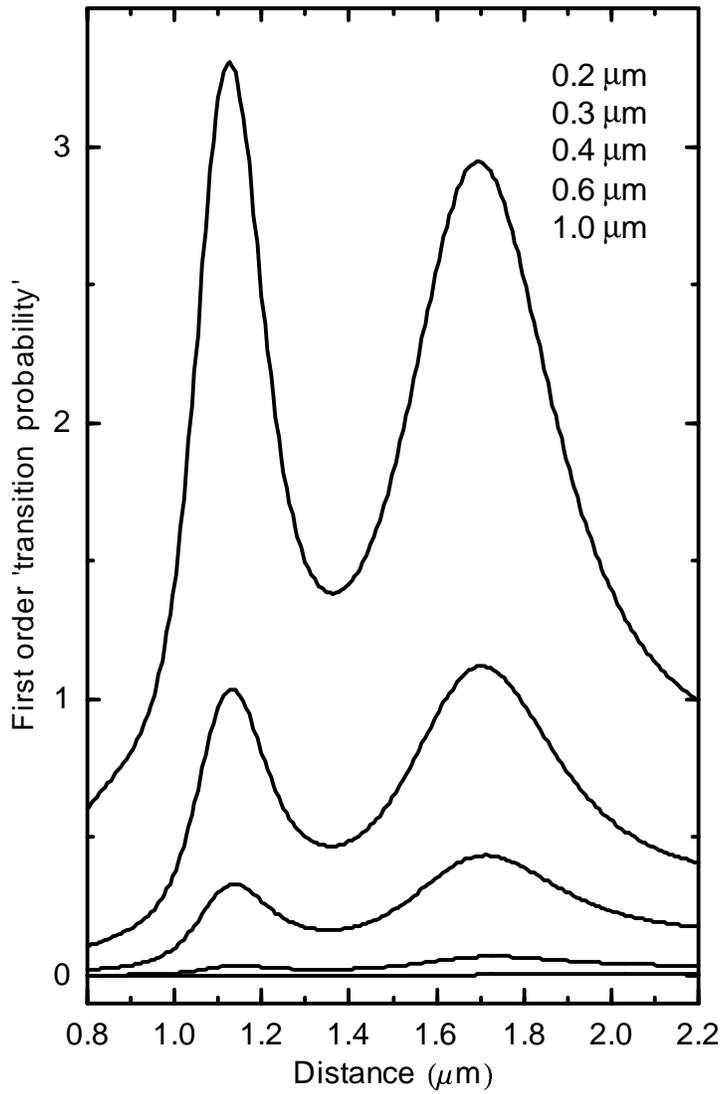
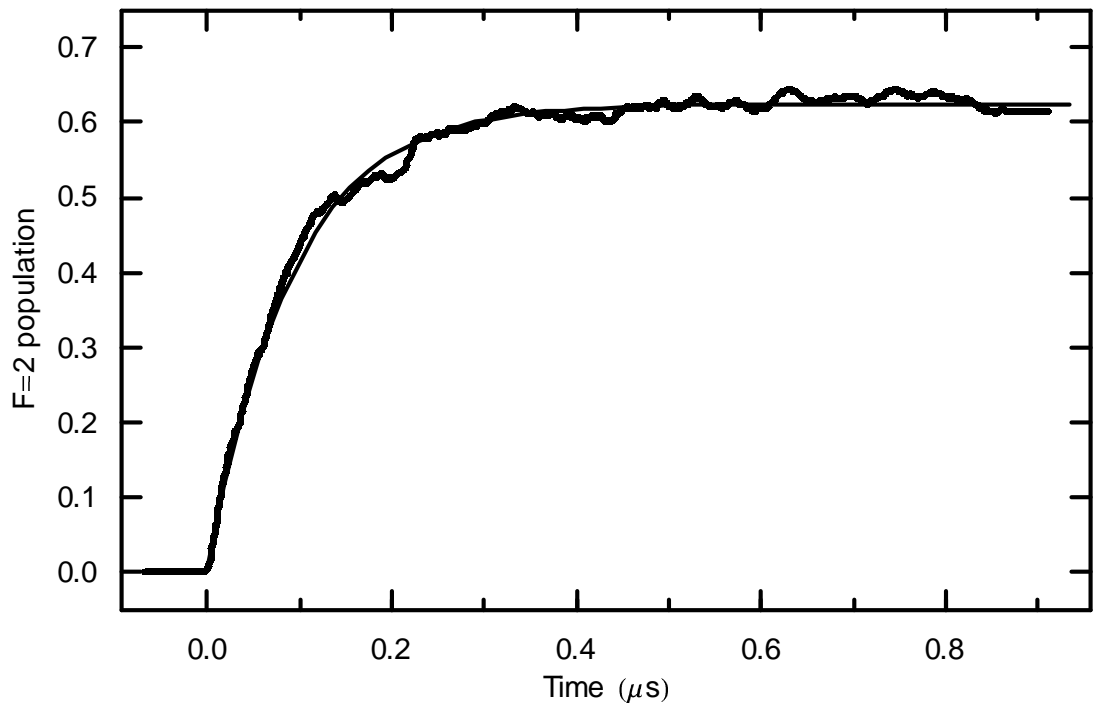
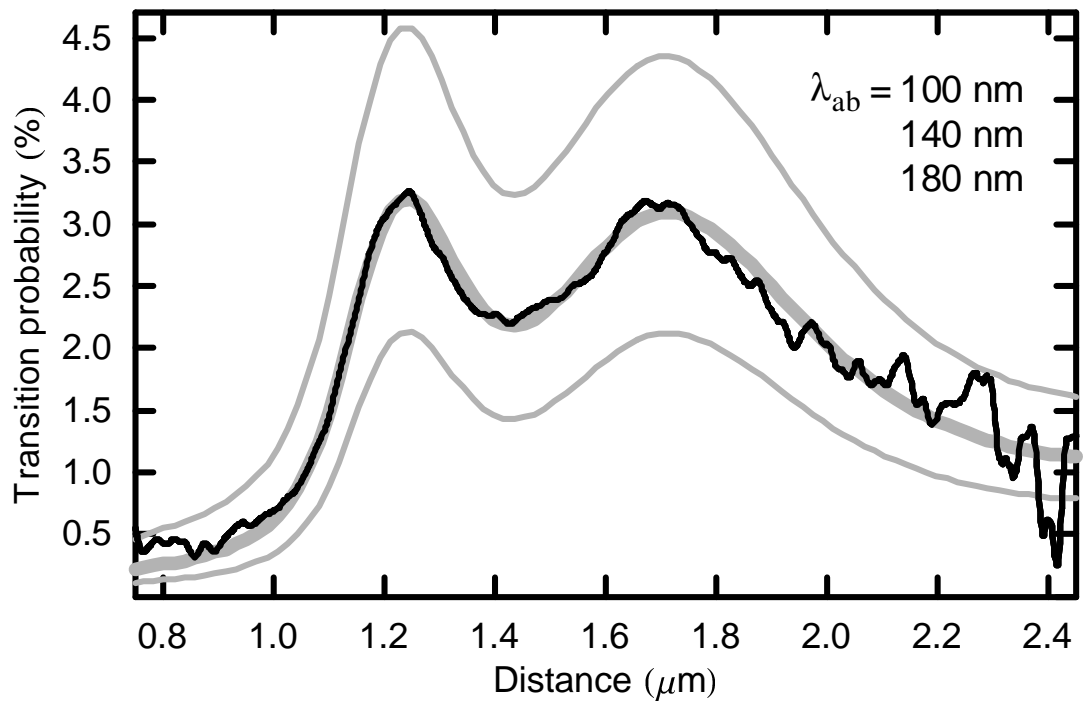
\end{document}